\documentclass[a4paper,notitlepage]{article}
\pdfoutput=1
\usepackage[english]{babel}
\usepackage{natbib}
\def\Po{\operatorname{Pois}}
\def\Ga{\operatorname{Gamma}}
\def\Beta{\operatorname{Beta}}
\def\Unif{\operatorname{U}}
\def\Bin{\operatorname{Bin}}

\def\Var{\operatorname{Var}}

\def\mean{\mu}
\def\TruePre{Y_{i}^\text{b}}
\def\truePre{y_{i}^\text{b}}
\def\ObsPre{Y_{i}^{\star\text{b}}}
\def\obsPre{y_{i}^{\star\text{b}}}
\def\TruePost{Y_{i}^\text{a}}
\def\truePost{y_{i}^\text{a}}
\def\ObsPost{Y_{i}^{\star\text{a}}}
\def\obsPost{y_{i}^{\star\text{a}}}
\def\muiPre{\mu^{\text{b}}_{i}}
\def\meanPre{\mu}
\def\deltaMean{\delta}

\newcommand{\dpois}[2]{\frac{{(#2)}^{#1}}{#1!}\exp\left(-{#2}\right)}
\newcommand{\dbin}[3]{{#2 \choose #1}{#3}^{#1}(1-{#3})^{#2-#1} }
\newcommand{\dGa}[4]{\frac{{#3}^{{#2}}}{\Gamma({#2})}{#1}^{{#2}-1}\exp\left(-{#4} {#1}\right)}

\newcommand{\dGaR}[3]{{#1}^{{#2}-1}\exp\left(-{#3} {#1}\right)}
\newcommand{\dBetaR}[3]{{#1}^{{#2}-1} (1-{#1})^{{#3}-1}}

\usepackage{amsmath}
\usepackage{MnSymbol}

\usepackage{numprint}
\usepackage{booktabs}
\usepackage{colortbl}
\definecolor{lightGray}{gray}{0.85}

\usepackage{lineno}  

\usepackage{url}

\usepackage{graphicx}
\title{\bf Hierarchical modelling of faecal egg counts to assess anthelmintic efficacy}
\author{Michaela Paul\\
{Institute of Mathematics, University of Zurich, 
  Zurich, Switzerland}\\
\texttt{michaela.paul@uzh.ch}
\and
{Paul R.~Torgerson}\\
{Section of Epidemiology, Vetsuisse Faculty, 
  University of Zurich, 
Zurich, Switzerland}
\and
{Johan H\"oglund}\\
Department of Biomedical Sciences and Veterinary Public Health, \\
  Swedish University of Agricultural Sciences,  
  Section for Parasitology, 
  Uppsala, Sweden
\and
{Reinhard Furrer}\\
{Institute of Mathematics, University of Zurich, 
  Zurich, Switzerland.}
}
\begin{document}
\maketitle

\begin{abstract}
Counting the number of parasite eggs in faecal samples is
a widely used diagnostic method to evaluate parasite burden.
Typically a sub-sample of the diluted faeces is examined 
for eggs. The resulting egg counts are multiplied by a specific
correction factor to estimate the mean parasite burden.
To detect anthelmintic resistance, the mean parasite burden 
from treated and untreated animals are compared. 
However, this standard method has some limitations. In
particular, the analysis of repeated samples may produce quite 
variable results as the sampling variability due to the
counting technique is ignored.
We propose a hierarchical model that takes this sampling
variability as well as between-animal variation into account.
Bayesian inference is done via Markov chain Monte Carlo
sampling. 
The performance of the hierarchical model is illustrated by
a re-analysis of faecal egg count data from a Swedish study
assessing the anthelmintic resistance of nematode parasite
in sheep. A simulation study shows that the hierarchical
model provides better classification of anthelmintic resistance 
compared to the standard method.

\end{abstract}

\section{Introduction}\label{sec:intro}

The extensive use of anthelmintic drugs to expel parasitic worms has lead to an
increasing problem of anthelmintic resistance (AR) particularly to gastrointestinal 
nematodes in every livestock host \citep{kaplan-2004}.  Reliable methods to 
assess and to detect changes in drug efficacy are thus essential for effective 
control of parasitic infections.  Various in-vivo and in-vitro methods have been 
developed to detect AR \citep[see review by][]{demeler2012}.  
The most widely used (in-vivo) method is the faecal egg count reduction test 
(FECRT).
It compares the number of parasite eggs in faecal samples taken before and after 
treatment \citep{mckenna1990}, or in samples of treated and untreated animals as 
control group \citep{coles-etal-1992,coles2006detection},
based on the assumption that the faecal egg counts (FECs) adequately reflect
the presence of adult worms in the gastrointestinal tract.
For sheep and goats, resistance is declared if the percentage reduction in mean 
FECs is less than 95\% and if the lower limit of a corresponding 95\% confidence 
interval (CI) for the reduction in means is below 90\%. If only one of the two 
criteria is met, resistance is suspected \citep{coles-etal-1992}. 
For other livestock, different thresholds have been suggested. 

\begin{figure}
   \centering
   \makebox{\includegraphics[width=.6\textwidth]{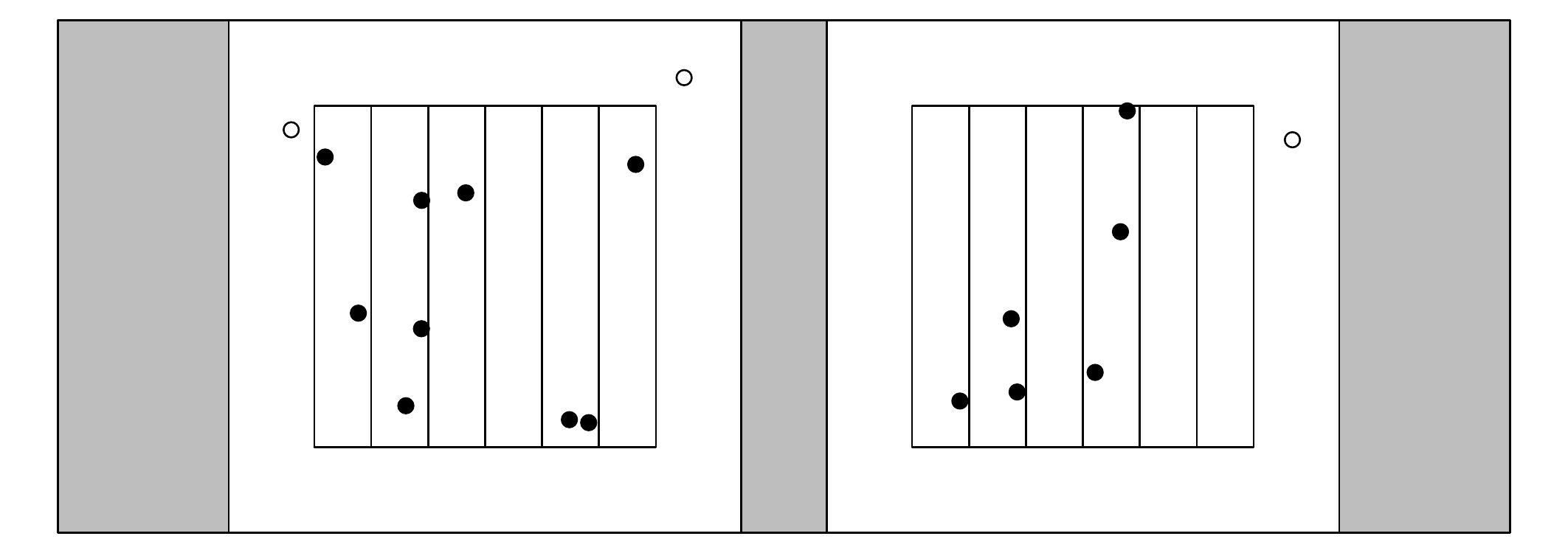}}
      \caption{\label{fig:McMaster}
	 Schematic representation of a McMaster slide. Eggs lying within the 
         two grid areas (filled circles) are counted to calculate 
         the number of eggs per gram of faeces. With a correction factor of 50, 
         this yields $(9+6)\times 50 = 750$ epg. }
\end{figure}

A major advantage of the FECRT is its easy and straightforward use 
for all three available broad-spectrum anthelmintic groups.
Standardised methods on how to obtain and compare FECs from various hosts
have been advocated by the World Association for the Advancement of Veterinary 
Parasitology (WAAVP) \citep{coles-etal-1992}. 
Typically, variants of the McMaster egg counting technique as detailed in
\citet{coles-etal-1992} are used to obtain FECs. 
First, the sample of faeces is diluted with a flotation fluid, thoroughly mixed 
and sieved to remove large debris. A sample of the resulting solution is then 
put in a counting chamber 
(see Figure \ref{fig:McMaster}). 
The eggs will float to the surface and can be counted.
Subsequently, the number of eggs per gram of faeces (epg) is calculated by 
multiplying the number of eggs observed within the grid areas by an appropriate 
factor which depends on the amount of faeces and flotation fluid used, as well 
as the total volume of the McMaster chambers ($2\times0.15$ml). Most commonly, 
faeces are diluted by a factor of 15, i.e.\ the method has an analytic sensitivity 
or detection level of $15/0.3=50$ epg.

The detection level influences the accuracy of the FECRT.
Results from repeated tests 
can be quite variable \citep{miller2006} particularly with low pre-treatment FECs, 
and are not reliable if only a small proportion ($\leq 25$\%) of the worms are resistant 
\citep{martin-1989}.
A reason for this is that the egg counting technique introduces (substantial)  
variability which usually is not accounted for \citep{torgerson-etal-2012}. 
Furthermore, the distribution of parasites and egg counts between animals is 
typically overdispersed \citep{grenfell1995}. 
Despite the limitations of the standard FECRT,
only few authors have proposed more elaborate statistical models for the analysis 
of overdispersed FECs. For instance, \cite{torgerson-etal-2005} assumed a negative 
binomial distribution for FECs from treated and untreated animals and used parametric
bootstrap to compute 95\% CIs for the reduction in mean.
\cite{denwood2010} assumed a Poisson-gamma model for the FECs, where 
the post-treatment means were linked to the pre-treatment means via scale factors. 
Inference was done via Markov chain Monte Carlo \citep[MCMC, e.g.,][]{Gilks1996} using JAGS software
\citep{plummer2003jags}.

To the best of our knowledge, no attempts have been made so far to also incorporate 
the variability of the egg counting process. In this paper, we propose a hierarchical 
model for the analysis of FECs that accounts for the above mentioned variabilities. 
The model is similar in spirit to econometric models for underreported counts 
which often occur in survey data due to varying reporting probabilities
\citep[e.g.][]{winkelmann2008,faderHardie-2000}.
Bayesian inference is done via tailored MCMC. 
In the following, we briefly review the conventional FECRT and summarise a
recent study that was undertaken to get an updated status of the prevalence
of AR of nematode parasites in Swedish sheep flocks \citep{hoeglund-etal-2009}.
We then discuss our hierarchical model formulation and computational approach for 
inference, followed by a re-analysis of the Swedish sheep FECs
and a simulation study comparing the ability of the FECRT and the hierarchical 
model to detect AR.
We close with some discussion.

\section{Faecal egg count reduction test} \label{sec:fecrt}
The WAAVP guidelines \citep{coles-etal-1992} detail a McMaster egg counting 
method with analytical sensitivity of 50 epg. To somewhat lessen the effect of the 
counting technique on the FECRT, 
they suggest to use groups of 
(at least) 10 animals, where the individual pre-treatment epgs should be 
$\geq$150. One group of $n_T$ animals is treated, and the other group
of $n_C$ animals serves as control. The percentage reduction is then calculated 
as
\begin{linenomath*}
\begin{align} \label{eq:fecrt}
  \text{FECRT}= 100\left(1- \frac{\bar{x}_T}{\bar{x}_C}\right),
\end{align}
\end{linenomath*}
where $\bar{x}_T$ and $\bar{x}_C$ denote the mean counts in the treated and 
control group, respectively.
Assuming independence, an approximate variance of the log ratio is given by
\begin{linenomath*}
\begin{align} \label{eq:var_fecr}
  \Var\left(\log \frac{\bar{x}_T}{\bar{x}_C}\right)= \frac{1}{\bar{x}_T^2}\Var(\bar{x}_T)+ \frac{1}{\bar{x}_C^2}\Var(\bar{x}_C).
\end{align}
\end{linenomath*}
This variance is used to construct a corresponding approximate 95\% CI on the 
log-scale using the 97.5\% quantile of a $t$-distribution with $n_T+n_C-2$ 
degrees of freedom, which is then backtransformed to give a 95\% CI
for the FECRT \eqref{eq:fecrt}.
For an anthelmintic to be fully effective, no worms should survive treatment 
and any viable eggs after treatment indicate that some worms may have been 
resistant. However, a 95\% or larger reduction in egg counts indicates that 
the anthelmintic treatment is still beneficial \citep{coles2006detection}.
As a decision rule, the WAAVP guidelines suggest that AR be considered present
in sheep and goats
if a) the reduction after treatment is less than 95\% \emph{and} 
b) the lower 95\% confidence limit is below 90\%.

Various (minor) variants of the FECRT and decision rules have been suggested 
in the literature depending on the study design and host species, see e.g.\
\citet{cabaret2004,miller2006}. For instance, without a distinct control 
group and faecal samples taken from the same animals before and after treatment,
\citet{mckenna1990} proposed to calculate egg count reduction as in \eqref{eq:fecrt}
with $\bar{x}_C$ now denoting pre-treatment counts.
A corresponding 95\% CI may be constructed using
(nonparametric) bootstrap \citep{Davison-Hinkley}. 
In practice, CIs for this paired situation are also constructed using variance \eqref{eq:var_fecr},
although this is not optimal from a statistical point of view.

\section{Anthelmintic resistance of nematode parasites in Swedish sheep flocks}
To assess the prevalence of anthelmintic resistance in Swedish sheep,
45 randomly selected farms with a 
minimum of 20 ewes were visited throughout Sweden 
during the grazing season of 2006 and 2007 \citep{hoeglund-etal-2009}. 
From each farm two flocks of approximately 15 lambs were dewormed with 
either benzimidazole (BZ) or macrocyclic lactone. In this paper, we only focus on the BZ treated 
flocks. Faecal samples were taken before treatment and analysed with a McMaster 
with analytic sensitivity of 50 epg. Out of the 45 flocks, 39 flocks with a mean 
of $\geq 50$ epg were re-sampled 7--10 days after treatment and also analysed 
with a McMaster. Flock sizes varied between 10 and 17 lambs and pre-treatment 
mean epgs were often low (ranging from 50 to 5580, with a median of 324 epg).

Mean egg count reduction was calculated using \eqref{eq:fecrt}. 
Both approximate 95\% CIs according to the WAAVP guidelines as well as
95\% bootstrap CIs were calculated to assess anthelmintic efficacy.
In 35 flocks, post-treatment FECs were all zero resulting in a FECRT of 100\%.
Of the remaining four flocks (IDs 24, 33, 36, and 39), BZ resistance was declared 
for two flocks (33 and 39).
Subsequent investigation of pooled pre- and post-treatment larval cultures showed
that \textit{Haemoncus contortus} was the main species involved in resistance.
BZ resistance of \textit{H. contortus} was further investigated with molecular 
tests of pre-treatment cultures. According to the molecular data, resistance was 
indicated by an estimated allele frequency of $\geq 95$\% in 
a total of five flocks (flocks 33 and 39 as already indicated by the FECRT, 
as well as flocks 24, 36, and 37).
In summary, \citet{hoeglund-etal-2009} concluded that the clinical resistance status 
of Swedish sheep nematodes is still
relatively low though resistance may be more widespread than indicated by the conventional FECRT.

\section{Modelling approach}\label{sec:model}
Suppose we have a group of $n$ untreated animals. A faecal sample from each 
animal $i$ is 
analysed by a McMaster method with 
analytic sensitivity $f$.
If the required amount of faeces is not available for certain animals, this 
results in larger correction factors $f_i$ for those animals. For notational 
simplicity we assume throughout the remainder of the text that the 
the same correction factor is used for all samples.
A fixed proportion $p=1/f$ of the diluted faecal suspension is put on
a McMaster slide and the eggs within the grid areas are counted.
Denote those observed raw counts by $\ObsPre$, $i=1,\ldots,n$,
where the superscript $b$ indicates that counts are collected before anthelmintic treatment.
Given the true number of eggs per gram of faeces, $\TruePre$, the observed number
of eggs $\ObsPre$ is binomial distributed with probability $p$ and size $\TruePre$.
The $\TruePre$ are assumed to be Poisson distributed with mean $\muiPre$.
Heterogeneity between animals is taken into account by letting $\muiPre$ 
follow a gamma distribution with shape parameter $\phi$ and rate parameter $\phi/\meanPre$ 
having mean $\meanPre$ and variance $\meanPre^2/\phi$.
This yields the following pre-treatment model
\begin{linenomath*}
\begin{equation}\label{eq:1sample}
\begin{split} 
   \ObsPre \,|\, \TruePre &\sim \Bin(\TruePre,p) \,, \\ 
   \TruePre \,|\, \muiPre &\sim \Po(\muiPre) \,, \\ 
   \muiPre \,|\, \phi,\meanPre &\sim \Ga(\phi,\phi/\meanPre) \,. 
\end{split}
\end{equation}
\end{linenomath*}

If the treatment is effective,
the number of eggs in faecal samples taken from the same animals 
some days after treatment should be vastly reduced. 
After treatment, the epg rate in \eqref{eq:1sample} is reduced by
a factor $\deltaMean$. This yields
\begin{linenomath*}
\begin{equation} \label{eq:2sample}
  \begin{aligned}
   \ObsPost \,|\, \TruePost &\sim \Bin(\TruePost,p) \,, \\
   \TruePost \,|\, \muiPre,\deltaMean &\sim \Po(\deltaMean\muiPre) \,, \\
  \end{aligned}
\end{equation}
\end{linenomath*}
where superscripts $a$ are now used to indicate the situation 
after anthelmintic treatment.
If the treatment is completely ineffective the underlying true epg rates after 
treatment should not change, whereas they should largely decrease in the 
case of an effective treatment. We thus assign a beta prior to the reduction 
parameter, $\deltaMean\sim \Beta(a_\delta,b_\delta)$.
Gamma priors are assigned to both the dispersion parameter and the population epg rate,
i.e.\ $\phi \sim \Ga(a_\phi,b_\phi)$ and
$\mean \sim \Ga(a_\mu,b_\mu)$.

Typical worm burden and FECs differ
depending on the animals (e.g.\ sheep, cattle) and types of parasites considered. 
Here we discuss the hyperparameters that will be used in the application 
in Section \ref{sec:application}.
The Swedish study investigated AR in sheep nematodes,
with the majority of eggs due to trichostrongylid infections.
Several authors have investigated the abundance and distribution of 
trichostrongylid eggs in sheep faeces.
For instance, \citet{grenfell1995} observed that overdispersion is 
correlated to the magnitude of the FEC and approaches a plateau for high mean 
egg counts. This is in line with the results by \citet{morgan-etal-2005} 
who investigated faeces of 14 groups of commercially farmed sheep.
Mean FECs ranged from 43 to 1915, and the estimated overdispersion parameter 
of a negative binomial distribution ranged from
0.18 (95\%-CI: 0.10--0.32) to 2.3 (95\%-CI: 0.2--4.2).

We thus assume a $\Ga(1, 0.7)$ 
as prior distribution for the overdispersion parameter $\phi$.
Then, 90\% of the prior probability mass lies between 
0.1 and
4.3
with a prior median of 
1.
For the population epg rate $\mean$ a more dispersed
$\Ga(1, 0.001)$ distribution is assumed,
where 90\% of the prior probability mass lies between 
51 and
2996.
Finally, we assign a $\Beta(1, 1)$
prior to the reduction parameter $\deltaMean$, so that
all values between 0 and 1 are equally likely a priori.
All priors are assumed to be (conditionally) independent.
The influence of the prior distribution for $\deltaMean$ on the results
is also investigated in a small sensitivity analysis.

\section{Markov chain Monte Carlo inference} \label{sec:mcmc}

Full Bayesian inference is based on the joint posterior distribution which is
proportional to 
\begin{linenomath*}
\begin{align*}
     &  \prod_{i=1}^n \left\{\dbin{\obsPre}{\truePre}{p} \;
                            \dbin{\obsPost}{\truePost}{p} \right. \\
    & \qquad\times \left. \dpois{\truePre}{\muiPre} \; 
                   \dpois{\truePost}{\deltaMean\muiPre} \;
                  \dGa{\muiPre}{\phi}{(\phi/\meanPre)}{\frac{\phi}{\meanPre}} \right\} \\
    & \qquad\times \dGaR{\meanPre}{a_\mu}{b_\mu}\;
                  \dGaR{\phi}{a_\phi}{b_\phi}\;
                  \dBetaR{\deltaMean}{a_\delta}{b_\delta}\,.
\end{align*}
\end{linenomath*}

Posterior marginal distributions for the parameters of interest cannot be
analytically derived, and we use MCMC sampling for inference.
If the full conditional distribution for a specific parameter 
is proportional to a known distribution Gibbs sampling \citep{casella-george-1992} 
is applied. 
Otherwise, samples from the full conditional distribution are obtained using
a Metropolis-Hastings (MH)  algorithm \citep{chib-greenberg-1995}
with a suitably chosen proposal distribution.

We start with the parameters for which Gibbs sampling is possible.
The full conditionals for the true number of epgs before treatment, $\obsPre$,
$i=1,\ldots,n$, are (displaced) Poisson distributions \citep{staff-1964} with mean 
$(1-p)\muiPre$ for $\truePre \geq \obsPre$ and zero probability for 
$\truePre = 0,1,\ldots,\obsPre-1$.
Similarly, the full conditionals for the true number of epgs after treatment, $\obsPost$,
$i=1,\ldots,n$, are (displaced) Poisson distributions with mean 
$(1-p)\deltaMean\muiPre$ for $\truePost \geq \obsPost$.
To update the individual epg rates before treatment, $\muiPre$, we draw samples
from a gamma distribution with shape $\truePre +\truePost + \phi$
and rate $\sum_{i=1}^n \muiPre + b_\delta$.

The full conditionals for the remaining three parameters have no analytically 
closed form. 
For updating the overdispersion parameter $\phi$ we draw samples from
the full conditional
\begin{linenomath*}
\begin{align*} 
   \displaystyle p(\phi\,|\,\cdot) \propto \frac{\phi^{n\phi +a_\phi-1}}{\Gamma(\phi)^n} \meanPre^{-n\phi}
	    \left(\prod_{i=1}^n \muiPre\right)^{\phi-1}
	    \exp\left(-\phi(1/\meanPre \sum_{i=1}^n \muiPre +b_\phi )\right)
\end{align*}
\end{linenomath*}
using a MH algorithm with uniform proposal distribution centred around the
current value (denoted by $\phi^{0}$) and suitably truncated to ensure that $\phi>0$,
i.e.\ proposal values are drawn from a 
$\Unif(\max\{0,\phi^{0}-s\},\, \phi^{0}+s) $ distribution 
(with tuning parameter $s$)
and accepted with probability 
$\min \left\{ 1,\; p(\phi^\star \,|\, \cdot)/p(\phi^{0}\,|\, \cdot)  
                  \times q(\phi^{0}\,|\, \phi^\star)/q(\phi^\star \,|\,\phi^{0} ) \right\}$.

For the population epg rate $\meanPre$ we have
\begin{linenomath*}
\begin{align}\label{eq:fc_mu}
\displaystyle
p(\meanPre\,|\,\cdot) &\propto \meanPre^{-(n\phi-a_\mu+1)} \exp\left( -b_\mu \meanPre -\frac{\phi\sum_{i=1}^n \muiPre}{\meanPre} \right) 
 = \exp\left(- a\log(\meanPre) -b\meanPre -\frac{c}{\meanPre}\right)
\end{align}
\end{linenomath*}
with $a=n\phi-a_\mu+1$, $b=b_\mu$, and $c=\phi\sum_{i=1}^n \muiPre$.
To sample from this full conditional distribution, we use a MH algorithm with
a suitable approximation of \eqref{eq:fc_mu} as proposal distribution.
The full conditional may be approximated by an inverse gamma, log-normal or gamma
distribution with respective parameters chosen to match the mode and the curvature 
at the mode of \eqref{eq:fc_mu}. Further details about these approximations 
are given in Appendix \ref{ap:approxFC_mu}.

To determine which approximation is best, we compute the Kullback-Leibler 
divergence \citep{kullback-leibler-1951} between the full conditional 
\eqref{eq:fc_mu} and the approximating inverse gamma/log-normal/gamma
distribution.
In general, an inverse gamma distribution provides the best approximation if $a$
is large (say, $a>2$). For smaller $a$, a log-normal distribution is often suitable.

Finally, the full conditional for the reduction in mean $\deltaMean$ is
\begin{linenomath*}
\begin{align} \label{eq:fc_delta}
   \displaystyle
 p(\deltaMean\,|\,\cdot) \propto \deltaMean^{\sum_{i=1}^n\truePost + a_\delta-1} (1-\deltaMean)^{b_\delta-1} \exp\left(- \deltaMean \sum_{i=1}^n \muiPre \right) \,.
\end{align}
\end{linenomath*}
As proposal distribution in a MH algorithm, we use a beta distribution with
parameters chosen to match the mode and curvature at the mode of the above full
conditional distribution, see 
Appendix \ref{ap:approxFC_delta}. 

The methods are implemented in an R package \texttt{eggCounts}
available from 
\url{http://www.math.uzh.ch/as} or
\url{http://cran.r-project.org/web/packages/eggCounts/}.

\section{Application to Swedish FECRT study}\label{sec:application}

In the following, the FECs of 39 sheep flocks by \citet{hoeglund-etal-2009} are
re-analysed with the hierarchical model discussed in Section \ref{sec:model}.
For 28 
out of the 575 
BZ treated animals no post-treatment FEC is available. Most of those animals 
had a pre-treatment FEC of zero or one.
For the analysis, all animals with missing post-treatment FECs were excluded.
In addition, one animal had a pre-treatment epg of 30 which is not possible with
a correction factor of 50. Here, 3 eggs floating outside the grid areas
of the McMaster slide were counted with a factor of 10.
As only the eggs within the grid areas should be counted, we set this FEC to 
zero.

For each flock, we obtained $\numprint{10000}$ posterior samples using a 
burn-in period of $\numprint{10000}$ iterations and a thinning of 
10. The tuning parameter for the update of $\phi$ was selected to 
achieve an acceptance rate between 30 and 40 per cent.
Acceptance rates for $\mean$ and $\deltaMean$ were generally high
with an average of 98.8\% (range: 92.8--100.0\%)
and 98.1\% (range: 97.4--100.0\%).
Standard output diagnostics provided by the R-package \texttt{coda} \citep{R-coda} 
were applied to check for convergence.

Table \ref{tab:FECRT} shows the result of the FECRT as discussed in Section 
\ref{sec:fecrt} for the five BZ treated flocks for which the molecular data 
indicated anthelmintic resistance. In addition to the approximate 95\%
confidence intervals based on \eqref{eq:var_fecr} 
we have also used a nonparametric bootstrap approach
with paired sampling 
implemented in the R package \texttt{boot} \citep{R-boot} 
to compute 95\% bootstrap percentile intervals
based on $1999$ bootstrap samples. 
As the former approach ignores that counts are taken from the same animals 
before and after treatment, the resulting approximate confidence intervals are 
wider than the respective bootstrap intervals.
The estimated percentage reductions in mean epg rate ($100(1-\deltaMean)$) 
obtained with the paired model are quite similar to the standard FECRT.
There is a clear indication of AR resistance for flocks 33 and 39.
Post-treatment FECs for flock 37 were all zero, and thus the percentage
reduction in mean epg rate is estimated as 100\% by the FECRT, and no 
confidence intervals can be computed.
As the pre-treatment mean epg was rather low for this flock, 
the sampling variability due to the McMaster method plays an important
role and leads to a relatively wide HPD interval for the reduction in mean.

\begin{table}
 \caption{
  \label{tab:FECRT}
  FECR results for the five BZ treated flocks for which the molecular data 
  indicated anthelmintic resistance. Shown are the results of the FECRT 
  (based on Eq.~\eqref{eq:fecrt}, together with 95\% approximate confidence 
  intervals and bootstrap percentile intervals), as well as the results of the 
  paired model (posterior median of $100(1-\deltaMean)$ and 95\% HPD interval).  
  Results are shaded in grey if resistance is declared, and printed in italics 
  if resistance is suspected based on the WAAVP guidelines.
} 
\centering
\fbox{%
\begin{tabular}{llllcll}
\multicolumn{1}{l}{\bfseries flock}&
\multicolumn{3}{c}{\bfseries FECRT}&
\multicolumn{1}{c}{\bfseries }&
\multicolumn{2}{c}{\bfseries hierarchical model}
\\ \cline{2-4} \cline{6-7}
\multicolumn{1}{l}{}&\multicolumn{1}{c}{FECR}&\multicolumn{1}{c}{approximate CI}&\multicolumn{1}{c}{bootstrap CI}&\multicolumn{1}{c}{}&\multicolumn{1}{c}{median}&\multicolumn{1}{c}{HPD interval}\\
\midrule
   24 & \cellcolor{white}    99.0 & \cellcolor{white}   [96.3, 99.8] & \cellcolor{white}   [97.5, 99.9] &     & \cellcolor{white}   99.0 & \cellcolor{white}   [98.5, 99.4]\\
   33 & \cellcolor{lightGray}    82.0 & \cellcolor{lightGray}   [65.3, 90.6] & \cellcolor{lightGray}   [71.3, 89.4] &     & \cellcolor{lightGray}   82.0 & \cellcolor{lightGray}   [77.7, 86.0]\\
   36 & \cellcolor{white}    97.5 & \cellcolor{white}   [90.6, 99.4] & \cellcolor{white}   [93.9, 100.0] &     & \cellcolor{white}   97.1 & \cellcolor{white}   [94.0, 99.3]\\
   37 & \cellcolor{white}   100.0 & \cellcolor{white}    & \cellcolor{white}    &     & \cellcolor{white} \slshape   97.7 & \cellcolor{white} \slshape   [90.0, 100.0]\\
   39 & \cellcolor{lightGray}    92.3 & \cellcolor{lightGray}   [62.9, 98.4] & \cellcolor{lightGray}   [83.8, 97.3] &     & \cellcolor{lightGray}   92.0 & \cellcolor{lightGray}   [88.9, 94.8]\\
\end{tabular}}
\end{table}

\begin{figure}[!h]
\centerline{\includegraphics[width=.7\textwidth]{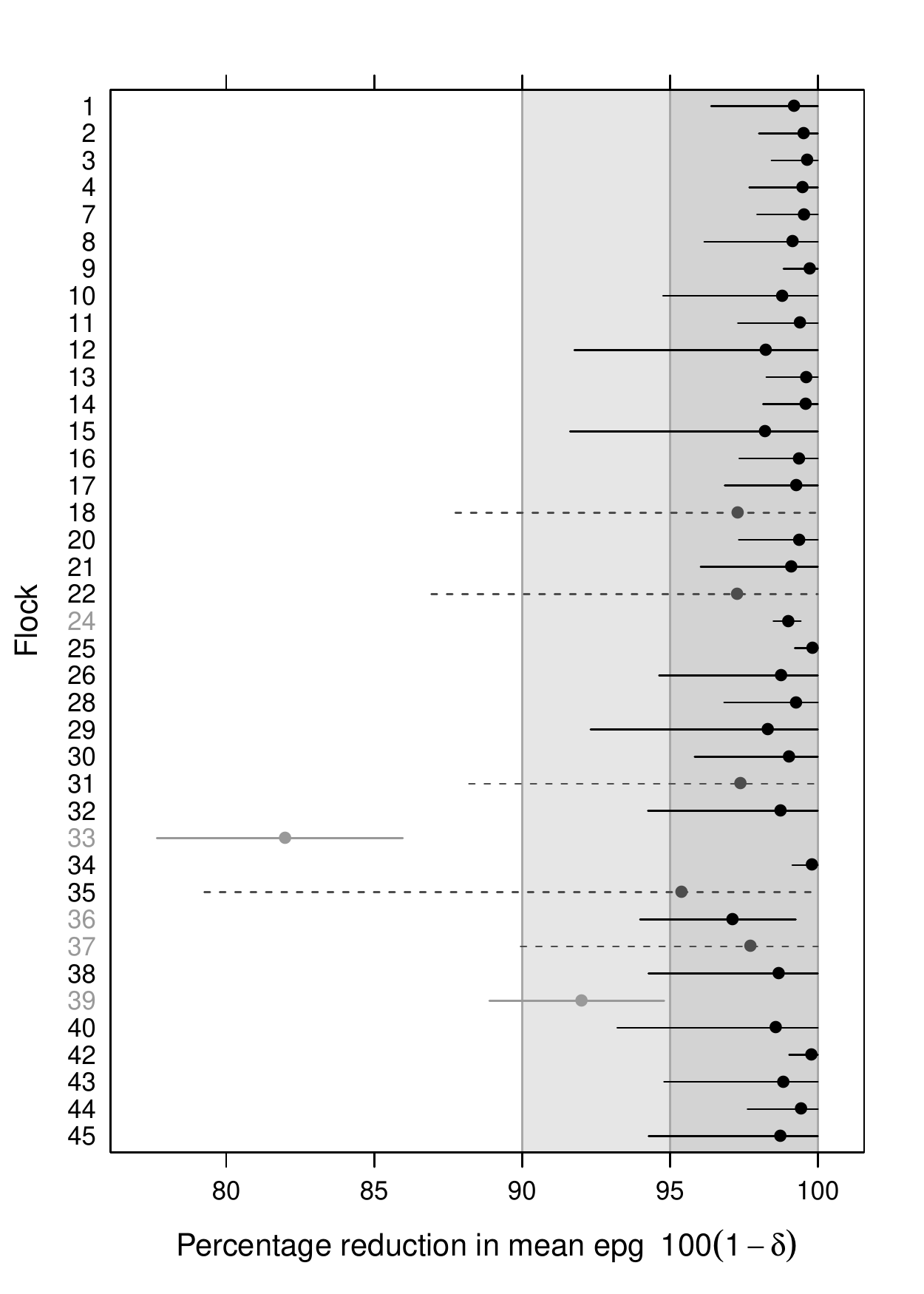}}
\caption{Estimated reduction in mean epg rate per flock treated with benzimidazole, 
   together with 95\% HPD intervals.
   Intervals are shown in grey if resistance is declared, and with dashed lines if resistance is 
   suspected based on the WAAVP guidelines.
   The flock number on the y-axis is printed in grey if anthelmintic resistance is suspected based on the
   molecular testing.
   \label{fig:SWE-fecr}} 
\end{figure}

Figure \ref{fig:SWE-fecr}
shows the estimated percentage reduction in mean epg
rate together with 95\% HPD intervals for all flocks, including the 35 flocks 
with zero post-treatment mean epgs for which the standard FECRT only provides 
an estimated reduction in mean of 100\% without confidence intervals.
The width of the intervals is largest for flocks 35 and 22 which had the lowest
pre-treatment means (below 100 epg).

\begin{figure}
\centerline{\includegraphics[width=.6\textwidth]{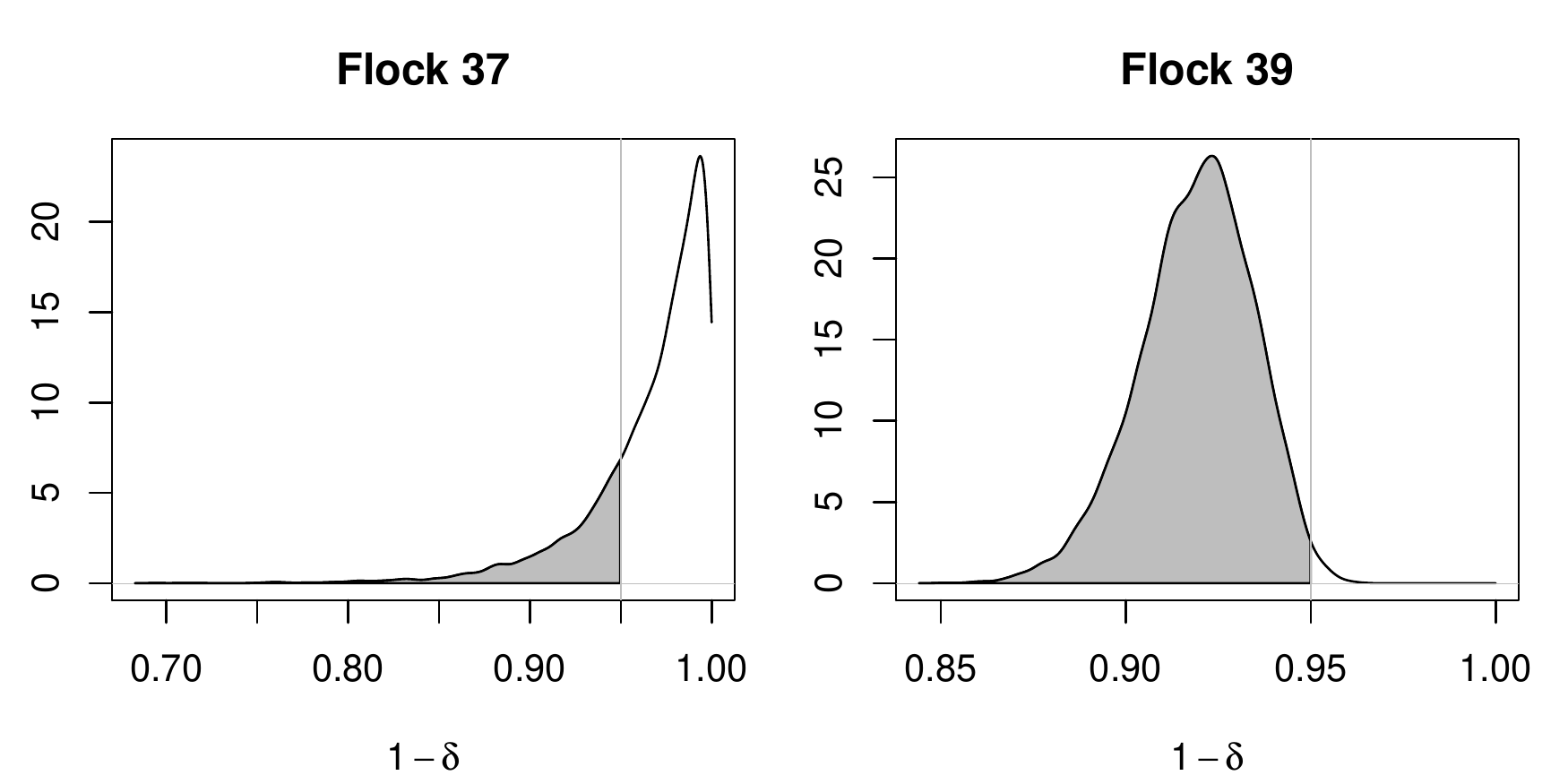}}
      \caption{\label{fig:postDelta}
	 Posterior marginal distributions for the reduction in mean ($1-\delta$) 
   for flocks 37 and 39. 
    The probability that the reduction in mean epgs is less than 95\% is shaded in grey.}
\end{figure}

An advantage of the Bayesian inference approach is that we obtain 
not only point estimates (and confidence intervals), but also
posterior marginal distributions for the parameters of interest.
For instance, we can make probability statements about the 
true reduction in mean being below or above some threshold.
These probabilities could then be used to judge AR \cite[see][]{denwood2010}. 
Figure \ref{fig:postDelta} shows the posterior marginals for 
the reduction in mean epgs $1-\delta$ for flocks 37 and 39.
According to the WAAVP guidelines for nematodes in sheep,
AR is present if the FECR is less than 95\%.
The posterior probability that the reduction in mean is less than 95\%
is shaded in grey in Figure \ref{fig:postDelta}.
For flock 37, this probability equals 0.22
indicating that there might be some resistance present.
In contrast, a probability of 0.99
for flock 39 provides strong evidence for AR.

We also investigated the sensitivity of the results to the prior assumptions.
In total, four different prior assumptions for $\deltaMean$ were considered:
\begin{enumerate}
   \item an uninformative Beta(1, 1) 
   distribution,
   \item an informative Beta(0.5, 1) 
   distribution with more weight on reductions close to 0 (i.e.\ no AR),
   \item an informative Beta(1, 0.5)
   distribution with more weight on reductions close to 1,
   \item a very informative Beta(5, 1)
   distribution with 99\% of the probability mass on values $\delta >$ 0.4
   (i.e.\ strong prior belief that AR is present). 
\end{enumerate}
The remaining model specifications stayed the same as before.

\begin{figure}[htb]
   \begin{center}
   \includegraphics[width=1\textwidth]{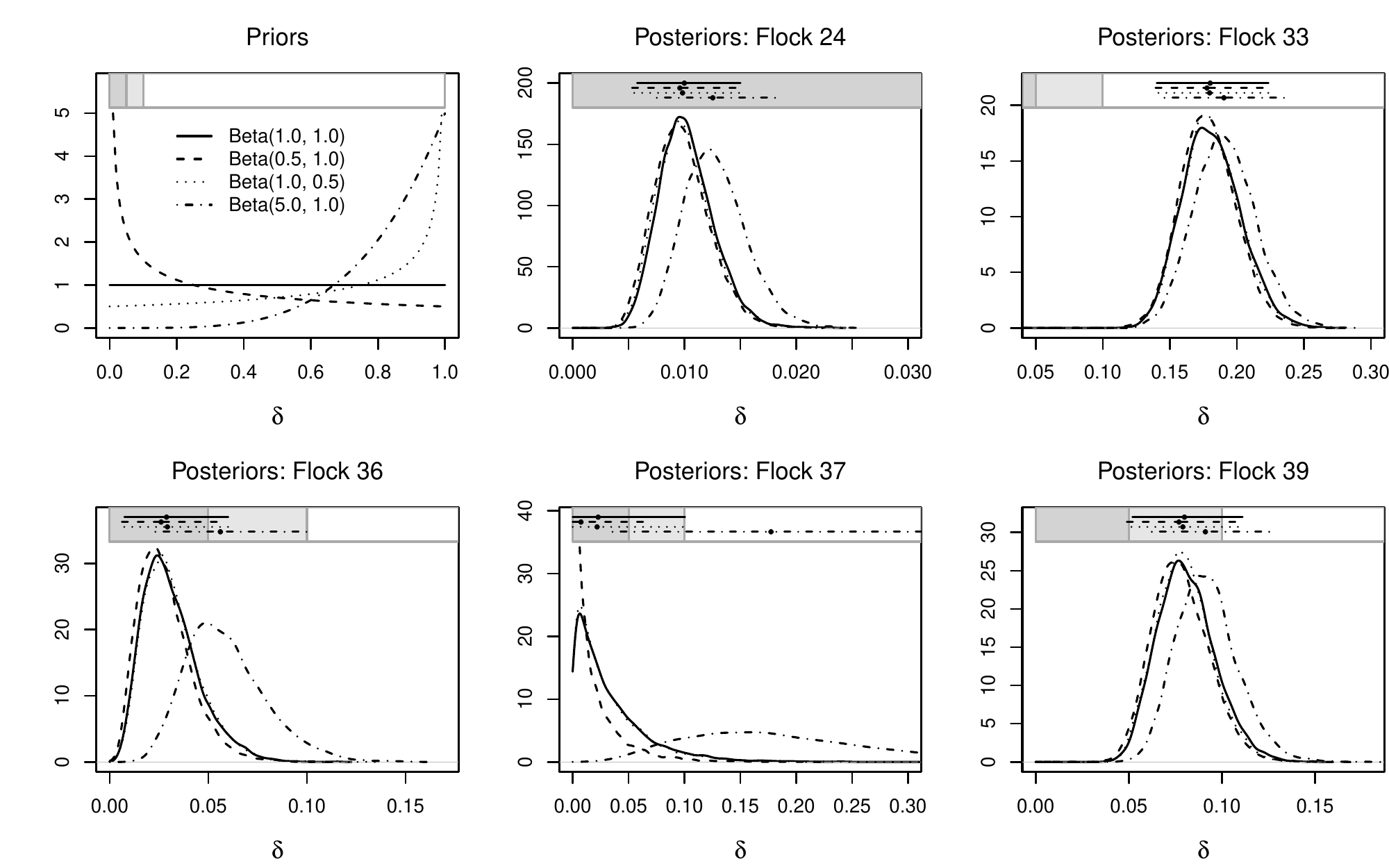}
      \caption{\label{fig:sensDelta}
   Sensitivity to the prior assumptions for reduction parameter $\delta$ 
   for the five BZ treated flocks in Table \ref{tab:FECRT}.
   The top left plot shows four different
   prior distributions for $\delta$. The remaining plots
   show the resulting posterior marginal distributions  
   together with respective 95\% HPD intervals in the top. 
   In addition, the definition of declared and suspected AR
   based on the WAAVP guidelines is illustrated by 
   the areas shaded in dark and light grey. 
	 }
   \end{center}
\end{figure}

Figure \ref{fig:sensDelta} shows the four prior distributions (top left) and
the resulting posterior marginal distributions for $\delta$ 
for the five BZ treated flocks for which the molecular data
indicated AR. 
In general, the posteriors are very similar for the first three priors,
and shifted somewhat to the right for the very informative prior.
In the top of the plots the respective 95\% HPD intervals are displayed.
Using the HPD intervals to decide on AR based on the WAAVP criteria
mostly leads to the same conclusions.
Results are most sensitive to the prior assumptions for flock 37 with 
very low pre-treatment mean FECs and zero post-treatment FECs.

\section{Simulation study}

To further investigate the ability of the hierarchical model to assess 
anthelmintic efficacy compared to the standard FECRT, a simulation study 
was conducted. We first simulated a number of pre- and post-treatment 
FECs for different anthelmintic efficacies and analysed them with both 
the FECRT and hierarchical model. We then computed the percentage 
of how often each method declared presence or absence of AR according to 
the WAAVP guidelines.

\subsection{Setup}
Similar to the Swedish FECRT study we consider flocks of $n=15$ 
animals. Pre-treatment FECs for these animals are simulated as follows:
We first draw mean epg rates $\mu_i$ from a gamma distribution with shape 
0.9 and rate $0.9/500$ and simulate the number of 
eggs $y_i^b$ from a Poisson distribution with this mean.
So marginally, the number of eggs are drawn from a negative binomial 
distribution with mean 500 and overdispersion parameter 0.9.
We then obtain the pre-treatment FECs (counted on a McMaster slide) by 
drawing from a Poisson distribution with mean $y_i^b/f$, where $f=50$ 
corresponds to the correction factor of the McMaster method.
Assuming an anthelmintic efficacy of $d$\%, the number of eggs after treatment 
$y_i^a$ are simulated from a Poisson distribution with mean $\mu_i(1-d/100)$. 
Post-treatment FECs are again obtained by sampling from a Poisson distribution 
with scaled mean $y_i^a/f$.

In total we considered 8 different scenarios with 
anthelmintic efficacies $d$ ranging from 85 to 99\%.
For each scenario, 2000 data sets were simulated and analysed with
the FECRT using a) an approximate CI ignoring the paired structure of the data, 
b) a paired bootstrap percentile interval, and 
c) the paired model \eqref{eq:1sample}--\eqref{eq:2sample}. For the latter, the same priors as 
discussed in Section \ref{sec:model} for the Swedish FECR study were used.
As before, we obtained $\numprint{10000}$ posterior samples using a 
burn-in period of $\numprint{10000}$ iterations and a thinning of 
10. The tuning parameter for the update of $\phi$ was selected to 
achieve acceptance rates between 30 and 40 per cent.

\subsection{Results}
\begin{figure}
   \begin{center}
   \includegraphics[width=1\textwidth]{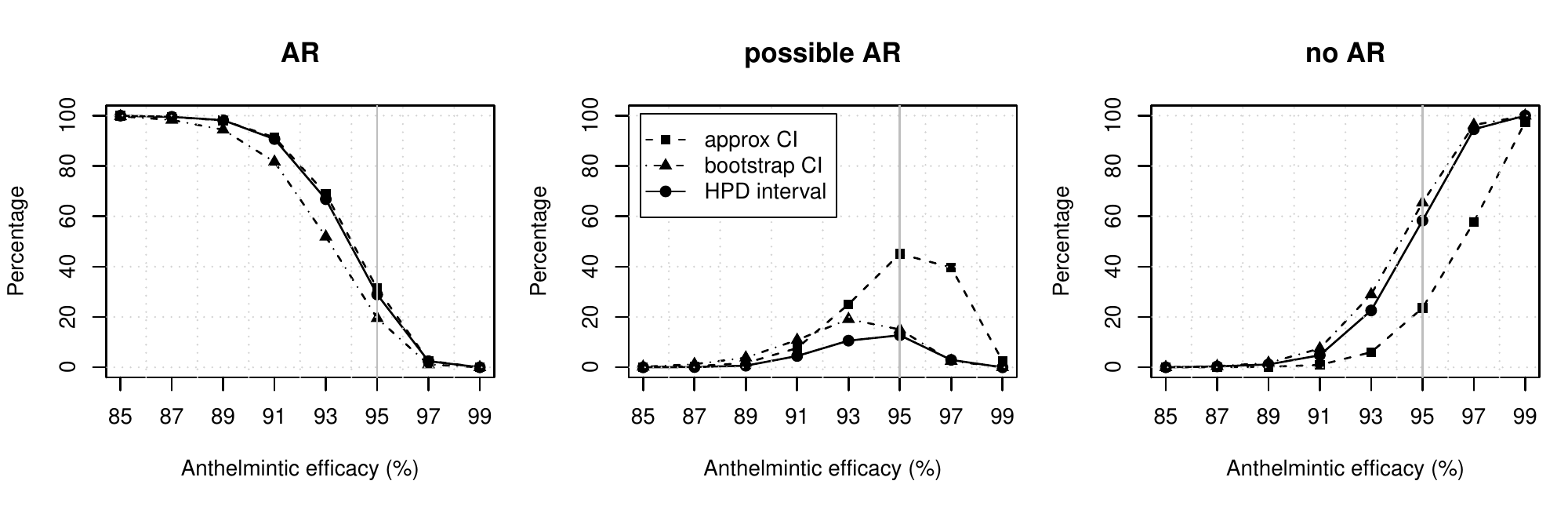}
      \caption{\label{fig:sim-probAR} 
  Results of the simulation study. Shown are percentages of how often the FECRT using an approximate CI ($\scriptscriptstyle\blacksquare$) or bootstrap percentile intervals ($\scriptstyle\blacktriangle$) and model \eqref{eq:2sample} (\textbullet) declared 
  anthelmintic resistance to be present (both criteria of the WAAVP guidelines are met), possible (one of the criteria is met), or absent (none of the criteria is met) for true anthelmintic efficacies ranging
  from 85 to 99\%.
	 }
   \end{center}
\end{figure}

The left plot in Figure \ref{fig:sim-probAR} shows how often the considered 
methods declared AR to be present based on the WAAVP guidelines, i.e. both the 
estimated reduction is less than 95\% and the lower limit of a 95\% confidence 
interval is less than 90\%.
Similarly the middle and right plot show how often AR is declared to be possible 
(one of the two criteria is met) and absent (none of the criteria is met). 
For a specific anthelmintic efficacy the percentages in the three plots sum to 
1 for each method. For the FECRT methods, AR is considered to be absent if the 
estimated reduction equals 100\% and no confidence interval can be computed due 
to a post-treatment variance of the FECs of 0. 
Suppose we consider the treatment as effective if the anthelmintic efficacy $d$ 
is larger than 95\%. This is indicated with a solid grey vertical line in Figure
\ref{fig:sim-probAR}.

For large values of $d=99$\% and for low efficacies ($d=85$--87\%), all methods 
yield the correct decision for nearly all simulated data sets. Closer to assumed 
efficacies of $d=95$\%, the decision gets more difficult.
In general, the hierarchical model performs better than the FECRT with 
bootstrap CIs and selects the true category of present (absent) AR when 
anthelmintic efficacies are smaller (larger) than 95\% with highest percentage.
However, the percentages of falsely deciding on absent AR when $d=91$ or $93\%$ 
are lowest for the standard FECRT with approximate CIs.
This is mainly due to the large width of the approximate CIs resulting
in high percentages for the middle category where a clear decision on 
absence or presence of AR is not possible.

\begin{figure}
   \begin{center}
   \includegraphics[width=.9\textwidth]{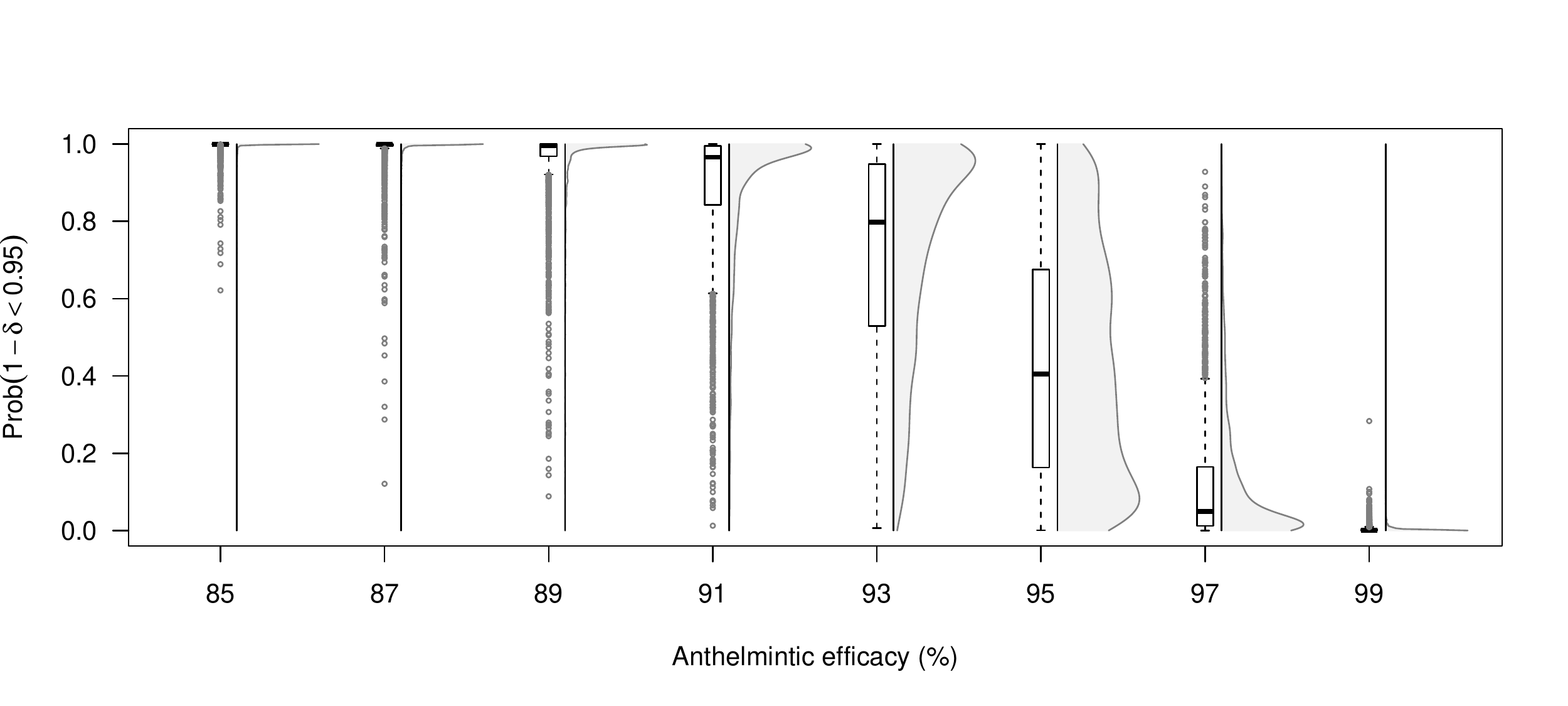}
      \caption{\label{fig:sim-p95}
   Results of the simulation study. Shown are boxplots and kernel density estimates of the posterior marginal probabilities that the reduction in mean epgs ($1-\delta$) is less than 95\% (indicating resistance) for true anthelmintic efficacies ranging
  from 85 to 99\%.
	 }
   \end{center}
\end{figure}

With the hierarchical model, an even better detection of AR can be obtained by 
using the posterior marginal distribution of $\delta$ for classification.
Figure \ref{fig:sim-p95} shows posterior marginal probabilities that the reduction 
in mean epg rates ($1-\delta$) is less than 95\%.
The same curve as for the HPD interval in Figure \ref{fig:sim-probAR} can be seen.
The further away from $d=95$\%, the more the probabilities are concentrated towards 
0 or 1. \cite{denwood2010} suggested that criteria 
$\text{Prob}(1-\delta < 0.95)> 0.975$  and 
$\text{Prob}(1-\delta < 0.95)< 0.025$
could be used to decide on 
``confirmed resistance'' and ``confirmed susceptibility''. 
With such a classification rule, no false decisions on absent AR occur
with efficacies $d<91\%$, and only 1 and 10 false decisions (out of 2000)
occur when $d=91$ and $93\%$, respectively.

\section{Discussion}
Anthelmintic resistance of nematodes in livestock is becoming more widespread.
The problem is serious in e.g.\ sheep nematodes and emerging e.g.\ in cattle nematodes.
Adequate methods to detect and monitor resistance are thus needed.
Most commonly FECs are used as a diagnostic tool to quantify parasite burden.
While many attempts have been made to improve the diagnostic procedures for
various hosts and parasites, the development of statistical methods for the 
analysis of FECs has received less attention.
In this paper we have proposed a hierarchical model that is able to incorporate 
sources of variability inherent in FECs.
In particular the variability resulting from the McMaster counting technique
is taken into account when estimating anthelmintic efficacy.
This is especially important in situations with low pre-treatment FECs and high
correction factors. Then the standard FECRT quite often results in an estimated
reduction in mean epg rate of 100\% without confidence interval.
In contrast, the hierarchical model provides a quantification of the uncertainty 
concerning the reduction parameter via the posterior marginal distribution.

For instance, in the application to the Swedish FECRT study
anthelmintic resistance could not be ruled out for flock 37
based on the results of the hierarchical model.
For this flock the molecular test also indicated AR. 
Note that the larval differentiation undertaken in this study provides
estimates of the abundance of the different nematode species. 
In some flocks, \textit{H.~contortus} was the prevailing species whereas
in other flocks \textit{H.~contortus} was only marginally present at the 
time of treatment with BZ.
If only a small proportion of eggs stems from the resistant species,
then it is very difficult to detect AR based on FECs.

An advantage of the hierarchical model formulation is its flexibility in model 
specification. In this paper we have assumed that the efficacy of the anthelmintic 
treatment is the same for each animal. This seems sensible as one would expect 
similar low efficacy in a resistant community. In a susceptible community efficacy 
should be high for all animals even though efficacy may be lower for some individuals
due to poor metabolism or availability of the drug in these animals \citep{cabaret2004}.
One could let the reduction parameter $\deltaMean$ vary across animals but
this would require adequate prior information about the efficacy distribution.

A further generalization of the model not considered here is the incorporation 
of zero-inflation. The WAAVP guidelines for sheep suggest to consider only 
animals with pre-treatment egg counts $>150$ epg \citep{coles2006detection}.
Hence, animals with zero pre-treatment FECs are sometimes discarded for the 
analysis. However, those counts do provide information about worm burden and 
should not be neglected. For instance, to address excess zeros (from uninfected 
animals) we can replace the gamma distribution for the individual epg rates by 
a mixture of a gamma distribution and a point mass at zero.

\section*{Acknowledgements}
The work of MP was supported by a grant from the Swiss National Science 
Foundation (ref: CR3313-132482). PT also recieved funding from the 
EU -- KBBE.2011.1.3-04 288975 ``Gloworm''. RF and MP acknowledge funding 
from the University Research Priority Program.

\appendix
\section{Proposal distributions} \label{ap:approxFC}
\subsection{Pre-treatment population mean $\meanPre$} \label{ap:approxFC_mu}

The unnormalized full conditional distribution of the pre-treatment
population mean epg rate $\meanPre$ is given in \eqref{eq:fc_mu} 
as
$ p(\meanPre\,|\,\cdot) \propto \exp\left(-G(\meanPre)\right)$
with $G(\meanPre) = a\log(\meanPre) +b\meanPre +c/\meanPre$.
The distribution has a mode at
\begin{linenomath*}
\begin{align} \label{eq:mode} 
\displaystyle m=\frac{-a +\sqrt{a^2+4bc}}{2b},
\end{align}
\end{linenomath*}
and the second derivative of minus the log density with respect to $\meanPre$ 
and evaluated at $m$ is given by 
\begin{linenomath*}
\begin{align}  \label{eq:D2} 
\displaystyle G''(m)=-\frac{a}{m^2} +\frac{2c}{m^3}.
\end{align}
\end{linenomath*}
We can approximate the full conditional by matching the mode \eqref{eq:mode} 
and the second derivative \eqref{eq:D2} to the respective values of a known 
distribution.

For instance, an inverse gamma distribution with shape parameter $\alpha$ and 
scale parameter $\beta$ has mode $m_\text{IG}=\beta/(\alpha+1)$, 
and second derivative of minus the log density at the mode equal to  
$D_\text{IG}=-(\alpha+1)/m_\text{IG}^2 +2\beta/m_\text{IG}^3$.
Replacing $m_\text{IG}$ and $D_\text{IG}$ by \eqref{eq:mode}--\eqref{eq:D2},
and solving the two equations for the shape and scale parameter yields
\begin{linenomath*}
\begin{align*}
   \alpha = \beta/m-1 \,,\qquad \beta = G''(m) m^3 
\end{align*}
\end{linenomath*}
as parameters of an inverse gamma proposal distribution for $\meanPre$.

Analogously, we can approximate the full conditional distribution
\eqref{eq:fc_mu} by a log-normal distribution with
mean $\log(m) + 1/G''(m)m^2$ and standard deviation $(\sqrt{G''(m)} m)^{-1}$,
or by a gamma distribution with shape $m\beta+1$ and rate $G''(m) m$.

\begin{figure}[tb] 
\centering
\includegraphics[width=1\textwidth]{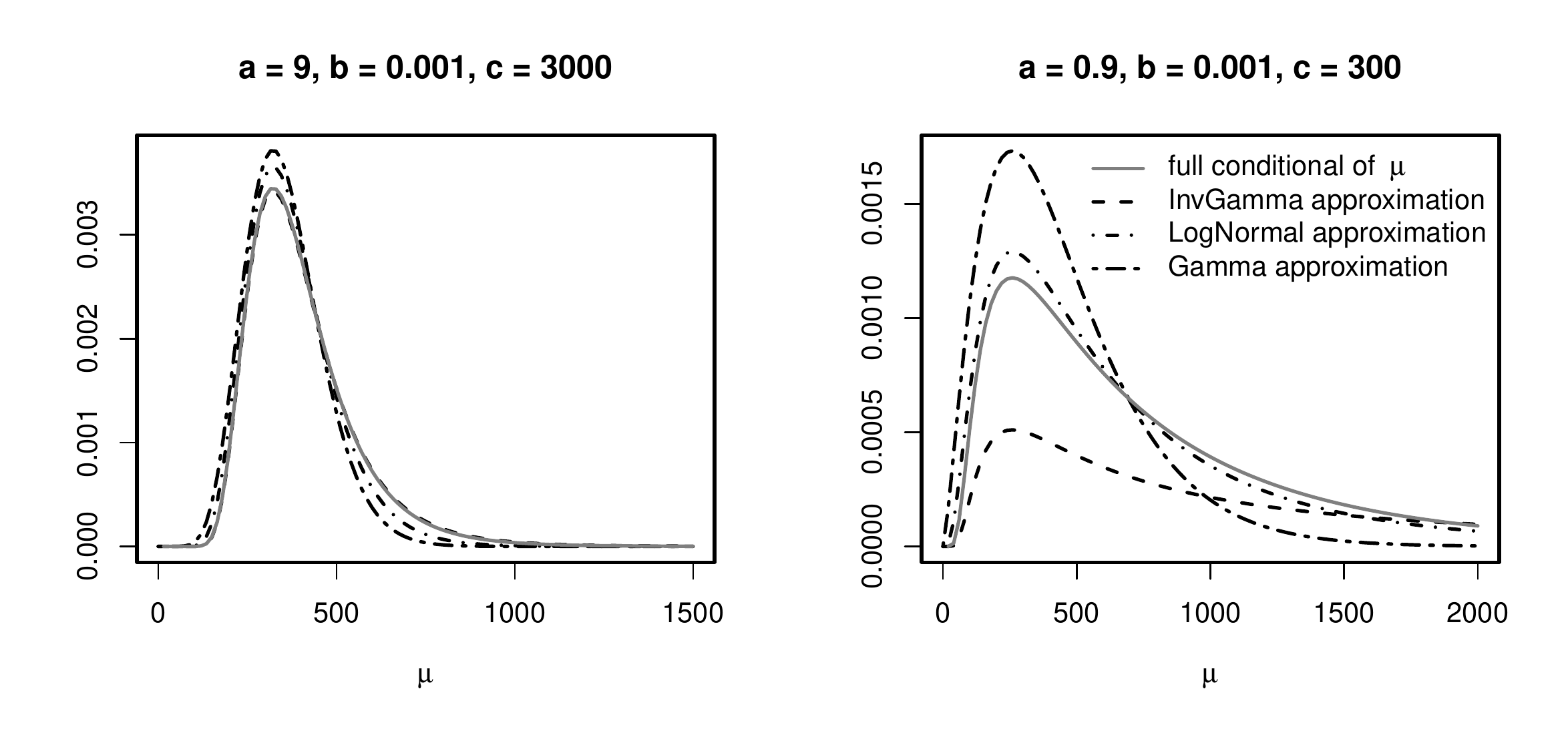}
\caption{Gamma, inverse gamma, and log-normal approximation of the full conditional 
  distribution of $\mean$ for two sets of parameters.
   \label{fig:approxFC}}
\end{figure}

Figure \ref{fig:approxFC} 
shows the full conditional together with the approximations
for two different sets of parameter values. 
For the first parameter set, the inverse gamma
distribution provides a very good approximation, whereas for the second set a
log-normal distribution is better suited.

\subsection{Reduction in mean $\deltaMean$} \label{ap:approxFC_delta}
The unnormalised full conditional distribution of $\deltaMean$ is given 
in \eqref{eq:fc_delta} as
\begin{linenomath*}
\begin{align*} 
   \displaystyle
 p(\deltaMean\,|\,\cdot) &\propto 
 \exp\Big(-[-a\log(\deltaMean)- b\log(1-\deltaMean) + c\deltaMean ] \Big) 
     = \exp\left(-G(\deltaMean)\right) 
\end{align*}
\end{linenomath*}
with $a=\sum_{i=1}^n\TruePost + a_\delta-1$, $b=b_\delta-1$, and $c= \sum_{i=1}^n \muiPre$.
The distribution has a mode at
\begin{linenomath*}
\begin{align} \label{eq:mode_delta} 
m= -\frac1{2c} (-a-b-c+\sqrt{a^2+2ab-2ca+b^2+2bc+c^2})
\end{align}
\end{linenomath*}
and the second derivative of minus the log density at the mode $m$ is given by
\begin{linenomath*}
\begin{align} \label{eq:D2_delta} 
G''(m) = \frac{a}{m^2}+\frac{b}{(1-m)^2}  \,.
\end{align}
\end{linenomath*}
Matching mode \eqref{eq:mode_delta} and second derivative \eqref{eq:D2_delta} to the
respective values of a $\Beta(\alpha, \beta)$ distribution yields
\begin{linenomath*}
\begin{align*}
   \alpha = (1-m)m^2G''(m) +1 \,,\qquad \beta = m(m-1)^2 G''(m) +1
\end{align*}
\end{linenomath*}
as parameters of an approximating beta distribution as proposal distribution 
for $\deltaMean$.

\bibliographystyle{chicago} 
\bibliography{references}

\end{document}